\documentclass[applmech,article,preprints,pdftex,moreauthors]{Definitions/mdpi}

\firstpage{1} 
\pubvolume{1}
\issuenum{1}
\articlenumber{0}
\pubyear{}
\copyrightyear{}
\datereceived{ } 
\daterevised{ } 
\dateaccepted{ } 
\datepublished{ } 
\hreflink{https://doi.org/} 

\usepackage{amssymb}
\usepackage{float}

\usepackage[utf8]{inputenc}
\usepackage[T1]{fontenc}
\usepackage{mathptmx}
\usepackage{etoolbox}
\usepackage{booktabs}
\usepackage{float}
\usepackage{multirow}
\usepackage[many]{tcolorbox}
\tcbuselibrary{skins}
\usepackage{tabularx}
\usepackage{array}
\usepackage{colortbl}

\usepackage{subcaption}
\usepackage[warn]{textcomp}

\tcbset{tab2/.style={enhanced,fonttitle=\bfseries,fontupper=\scriptsize\sffamily,
colback=white!10!white,colframe=red!50!black,colbacktitle=\color{Blue!40!white},
coltitle=black,center title}}

\definecolor{blue}{rgb}{0.0, 0.0, 1.0}

\tcbset{tab1/.style={enhanced,fonttitle=\bfseries,fontupper=\scriptsize\sffamily,
colback=white!10!white,colframe=blue!50!black,colbacktitle=blue!20!white,
coltitle=black,center title}}

\Title{Parametric Study of a Bladeless Fan Geometry: Investigating the Influence of Geometry Parameters on Discharge Ratio and Thrust Force}

\TitleCitation{Title}

\Author{
	Guillaume Maîtrejean $^{1,\dagger,\ddagger}$*\orcidA{},
	Merlin Kempf $^{1,\ddagger}$,
	Lucas Antoniali $^{1,\ddagger}$,
	Feirrera-Lopes $^{1,\ddagger}$,
	Mohamed Karrouch $^{1,\ddagger}$,
	Didier Bleses $^{1,\ddagger}$,
	François Truong $^{1,\ddagger}$,
	Denis C.D. Roux $^{1,\ddagger}$
	}

\AuthorNames{Firstname Lastname, Firstname Lastname and Firstname Lastname}

\address{%
$^{1}$ \quad {Univ. Grenoble Alpes, CNRS, Grenoble INP, LRP},
{Grenoble},
{38000}, 
{France}}

\corres{Correspondence: guillaume.maitrejean@univ-grenoble-alpes.fr}

\abstract{The innovative bladeless design of the AirMultiplier, patented by Dyson in 2009 \cite{gammack2009bladeless}, has generated significant interest due to its unique approach to air circulation that leverages the Coanda effect. In this study, we present a comprehensive parametric analysis of a simplified bladeless fan geometry, with a specific focus on the discharge ratio and generated thrust force for small to medium radii (5 to $200mm$) and a range of flow rates spanning from $1$ to $100g.s^{-1}$. In addition to the radii and mass flow rates, this study also addresses the effect of the slit nozzle thickness on the discharge ratio and thrust force of the AirMultiplier. By exploring the influence of this additional parameter, we aim to provide a more complete understanding of the performance characteristics of the geometry and to offer insights that could inform the design and optimization of similar bladeless geometries.}


\begin{document}


\section{Introduction }
\label{sec:intro}
\begin{figure}[H]
        \centering
        \includegraphics[width=5cm]{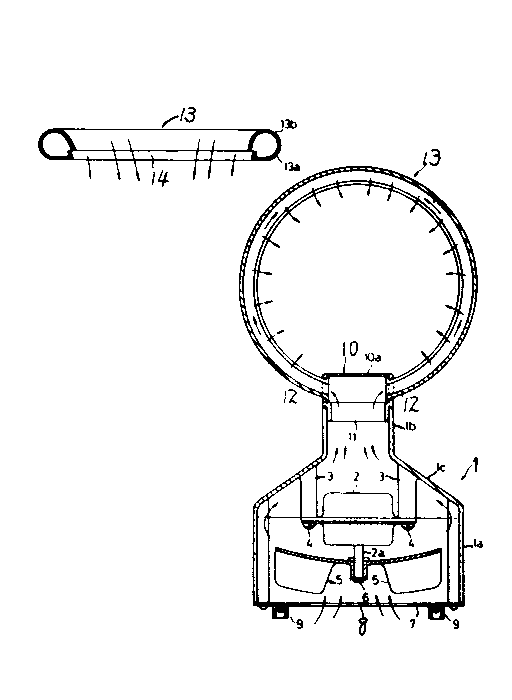}
        \caption{Schematic of the Toshiba bladeless fan from \cite{toshibabladeless}.}
        \label{fig:tosh}        
\end{figure}

In 1981, Toshiba first patented \cite{toshibabladeless} a bladeless fan design consisting of a ring with a slit as illustrated Fig. \ref{fig:tosh}.

\begin{figure}[H]
        \centering
        \includegraphics[width=5cm]{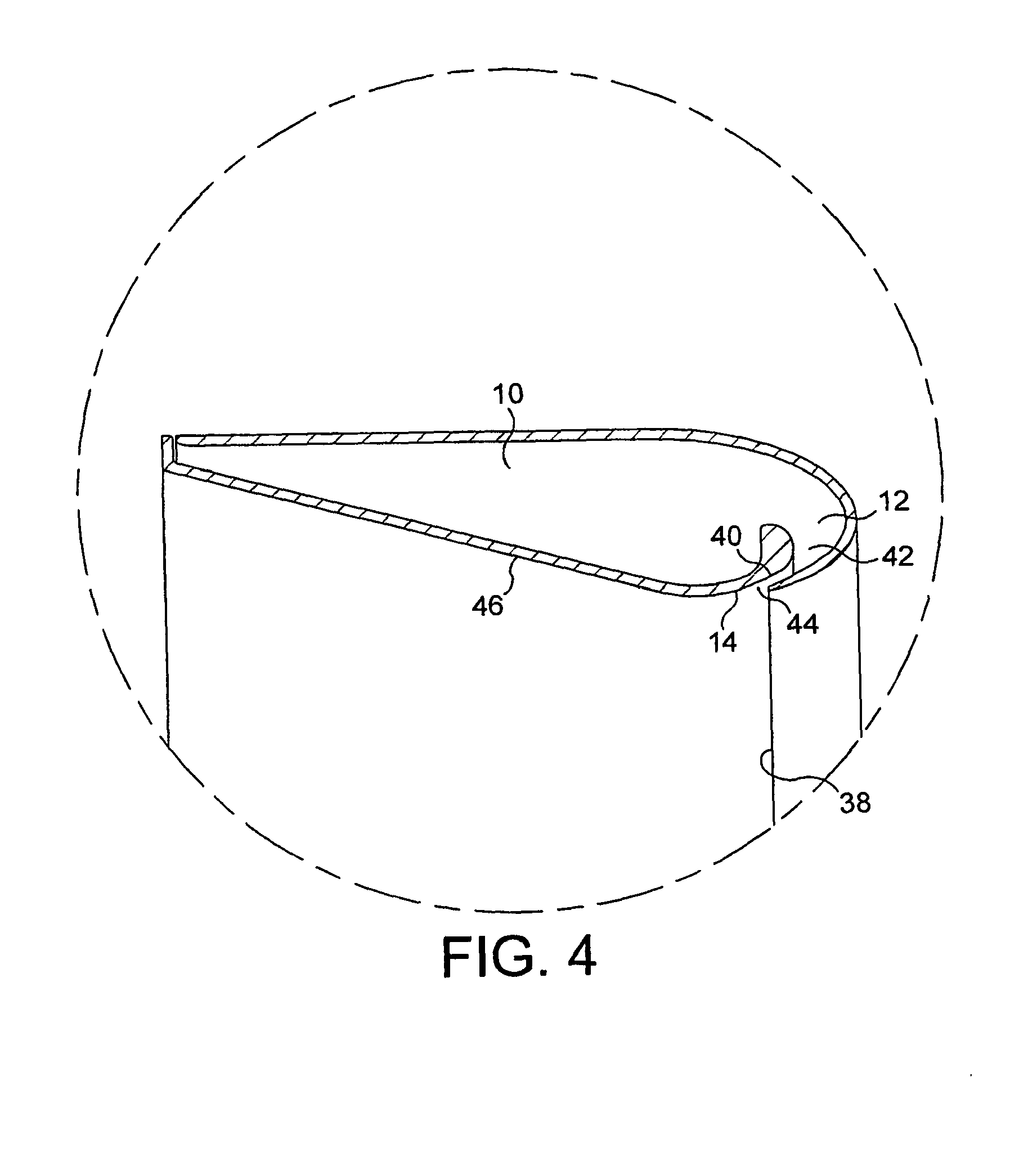}
        \caption{Schematic section of the Dyson bladeless fan from \cite{gammack2009bladeless}.}       
        \label{fig:dyson} 
\end{figure}

This first geometry was later improved upon by Dyson in \cite{gammack2009bladeless}, who incorporated the Coanda effect, which describes the tendency of a fluid jet to adhere to a nearby surface, causing it to curve and follow the contour of the surface, by using a reverse engineered aircraft wing geometry (see Fig. \ref{fig:dyson}). 

In the case of the bladeless fan, a thin, annular jet of air is expelled through a narrow slit nozzle. As the high-velocity air stream exits the nozzle, it adheres to the curved surface of the fan's housing, creating a vortex-like flow pattern.

Simultaneously, the Bernoulli principle comes into play. According to this principle, an increase in the speed of a fluid is accompanied by a decrease in pressure. As the high-speed air stream passes over the curved surface of the fan, its velocity increases, resulting in a decrease in pressure. This pressure differential between the inner and outer regions of the fan creates a low-pressure zone on the inner side. Air from the surroundings is drawn into this low-pressure zone and entrained into the airflow, amplifying the volume of air being circulated. \newline

\subsection*{Discharge ratio}
The performance of the bladeless fan is often quantified by the discharge ratio, which is the ratio of the mass flow rate of air expelled by the fan to the mass flow rate of air drawn into the fan from the surroundings. This discharge ratio is a measure of the efficiency and effectiveness of the fan in generating airflow.

\subsection*{Thrust force}
In addition to the airflow, the bladeless fan also generates a (generally small) thrust force. The thrust force is the force exerted by the expelled air, propelling the fan forward. This thrust force, denoted by $F$, can be calculated using the equation:

\begin{equation}
    F = \dot{m} \cdot v_e
\end{equation}

where $\dot{m}$ is the total mass flow rate and $v_e$ is the air velocity at the housing outlet.

By optimizing the geometric parameters and design of the fan such as the nozzle slit thickness and the radius, the bladeless fan can generate significant thrust force.

\begin{figure}[H]
        \centering
        \includegraphics[width=15cm]{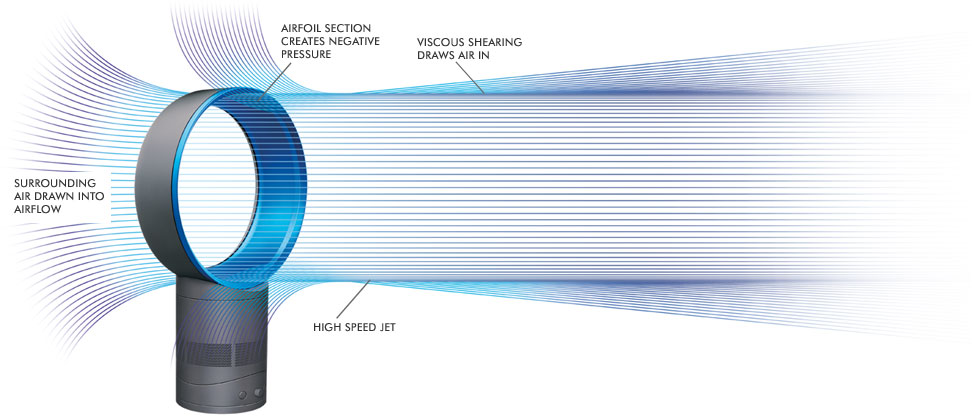}
        \caption{Illustration of the Dyson bladeless fan from \cite{dyson}.}       
        \label{fig:dysonSite}   
\end{figure}

Finally combining the Coanda effect and the Bernoulli principle, the bladeless fan creates a continuous and efficient flow of air without the need for traditional rotating blades as illustrated Fig.\ref{fig:dysonSite}. The curved surface of the fan's housing helps direct and control the airflow, while the pressure differential created by the high-speed air stream and the entrained surrounding air produces a powerful and focused airflow.

\subsection*{Geometric performance analysis}
The bladeless design of the Dyson AirMultiplier has since garnered significant attention due to its unique approach to air circulation, which eliminates the need for conventional rotating blades. Some studies have examined the influence some parameters such as shape on the outlet ﬂow ﬁeld structure \cite{li2016experimental}, discharge ratio \cite{joshi2023determination,mehmood2023design, jafari2016numerical, ravi2022numerical} or even noise \cite{jafari2014numerical}.

However, these studies address large radius geometry and have been limited in terms of their range of parameters and lack a systematic approach.

In light of these limitations, the present study aims to provide a more comprehensive and systematic investigation of the performance characteristics of the Dyson AirMultiplier geometry. Specifically, we focus on the effect of various geometric parameters, including the nozzle slit thickness on the discharge ratio and thrust force for small to medium radii (5 to 200mm) and a range of flow rates spanning from $1$ to $100g.s^{-1}$. By examining the influence of these parameters on the performance of the AirMultiplier, we aim to provide insights that can inform the design and optimization of similar bladeless fans, as well as advance our understanding of the underlying physics of this innovative device.
  
To perform this investigation, the numerical simulation tool Ansys Fluent is used, which allows us to model the flow field within the AirMultiplier and compute the corresponding discharge ratio and thrust force. We first describe the numerical model and simulation setup used in this study, including the geometry and meshing of the AirMultiplier, as well as the boundary conditions and solver settings used in the simulations.

Next, we present the results of our parametric study, which reveals a non-linear relationship between the mass flow rate discharge ratio and small radii and with an optimal radius for maximum generated thrust force. We also observe that the slit nozzle thickness and radius combinations can have a considerable effect on the generated thrust force, with some configurations able to generate the same amount of force as a De Laval nozzle.

Finally, we discuss the implications of these findings for the design and optimization of bladeless fans, as well as potential future research directions. Overall, this study provides a more comprehensive understanding of the performance characteristics of the bladeless fan geometry with small to moderate radii and a large range of both slit nozzle thicknesses and inlet mass flow rates, and highlights the importance of considering a range of geometric parameters in the design and optimization of similar bladeless fans.

\begin{table}[H]
        \centering
        \setlength{\tabcolsep}{2pt}
        \begin{tcolorbox}[tab1,tabularx*={\renewcommand{\arraystretch}{1.25}}{l|r},title={Parameters studied},boxrule=0.8pt, width=8cm]
        \hline
        Parameters &  Value / Range \\
        \hline
        Mass Flow Inlet (MFRi)  & 1, 10 and 100$fg.s^{-1}$ \\
        Slit thickness &  $0.25mm$ to $1.5mm$ \\
        Radius & $5mm$ to $200mm$ \\ 
        \hline
        \end{tcolorbox}
        \caption{Parameters investigated in the current study for the bladeless fan geometry.}
        \label{tab:parameters}
\end{table}

\section{Numerical Model}
\subsection{Geometry}
The bladeless fan geometry studied in the present article is a simplified model of the Dyson AirMultiplier geometry, which consists of a circular base with a centrally located cylindrical cavity. The cavity is connected to a slit nozzle, referred hereafter as \textit{slit}, that runs circumferentially around the outer edge of the cavity, creating a thin, annular jet of air. The slit thickness is assumed to be uniform and is fixed in this simplified model.

The cylindrical cavity is assumed to have a constant radius throughout its length, and the outer diameter of the circular base is also fixed. 

\begin{figure}[H]
        \centering
        \includegraphics[width=14cm]{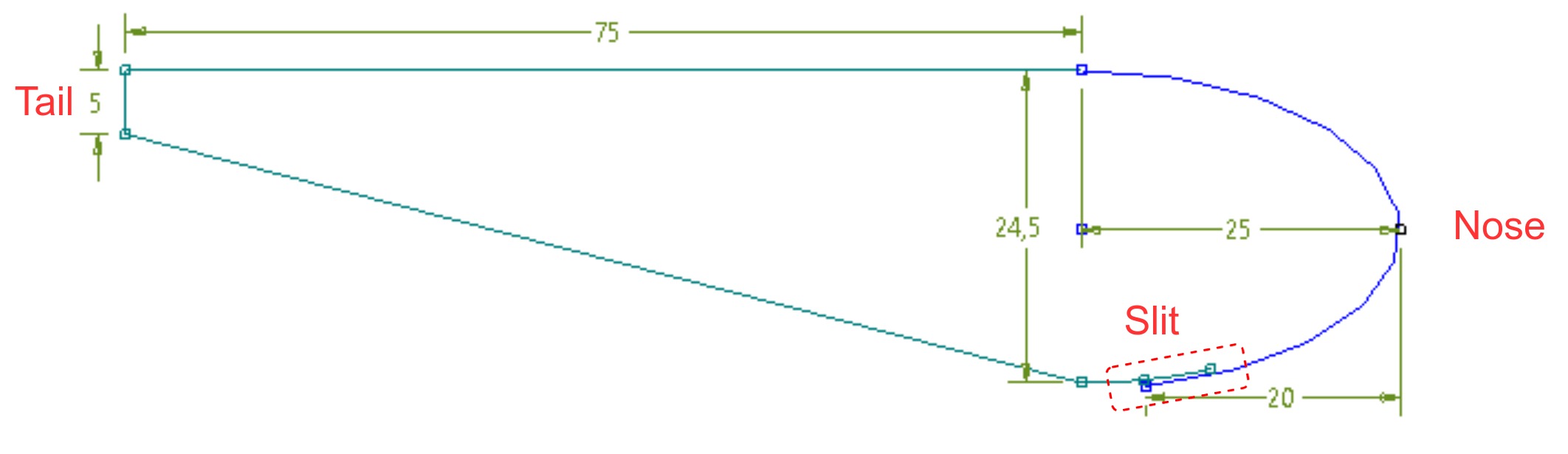}
        \caption{Section of the bladeless fan geometry used in the present article with a tail thickness of $5mm$ and the slit outlet located at $20mm$ from the nose. All dimensions specified in millimeters.}       
        \label{fig:geom} 
\end{figure}

The geometry of the bladeless fan in our study is simplified to focus on the influence of key geometric parameters such as the radius of the cavity and the nozzle/slit position and thickness on the discharge ratio and thrust force. The dimension of the geometry are detailed in milliter Fig.\ref{fig:geom}

To simplify the reading of the article and to avoid repetition, we will refer to the geometry either as a BladeLess Fan or \textit{BLF} geometry throughout the rest of the article and the radius will be understood as the minimum distance between the symmetry axis and the BLF.
Additionally, as illustrated in Fig. \ref{fig:geom}, the front of the geometry is called the nose and the back is referred to as the tail or outlet and the inlet mass flow rate is referred to as \textit{MFRi}.

\subsection{Governing Equations and Boundary Conditions}
\begin{figure}[H]
        \centering
        \includegraphics[width=14cm]{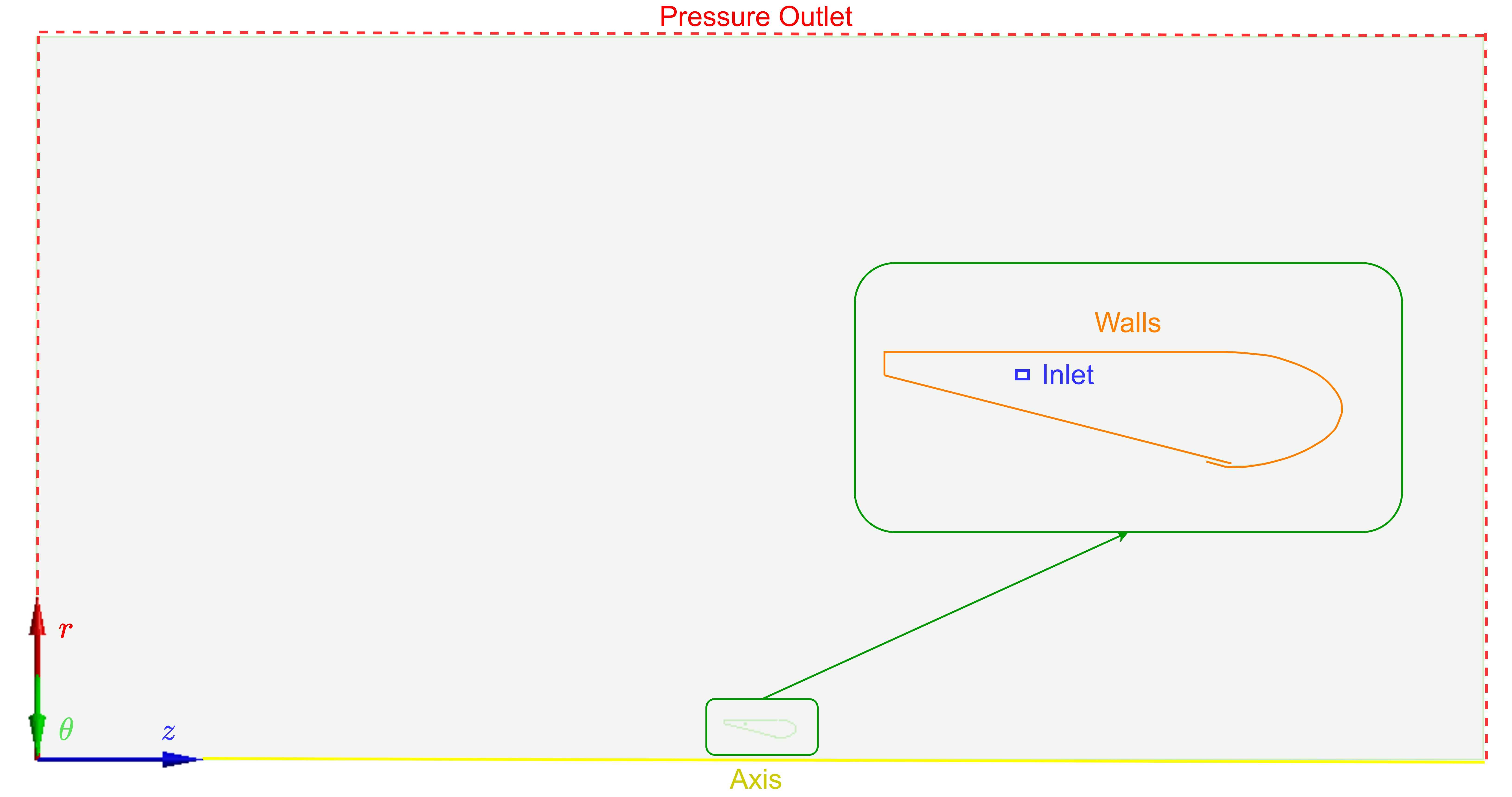}
        \caption{Simulation domain of the axisymmetric geometry with the thruster positioned at its center with the cylindrical coordinate systems associated to it.}       
        \label{fig:domain} 
\end{figure}

The model has been simulated using Ansys Fluent with the steady RANS Spallart-Allmaras turbulent model \cite{spalart1992one}. 
The fluid considered in the present study is air, assumed to be compressible and treated as an ideal gas. As the geometry is assumed to be axisymmetric, only a section of the domain needs to be simulated. The simulated domain section has a width of one meter and a height of 500mm, with the geometry placed at its center (see Fig \ref{fig:domain}).

The inlet boundary condition is set to a fixed mass flow rate (from 1 to $100g.s^{-1}$). The outlet boundary condition is set to a relative pressure of 0Pa. The walls are assumed to be adiabatic and the symmetry axis is located at $r=0$ in the cylindrical coordinate system associated to the geometry, as depicted on Fig.\ref{fig:domain}.

\subsection{Mesh and Model Validation}
\begin{figure}[H]
        \centering
        \includegraphics[width=14cm]{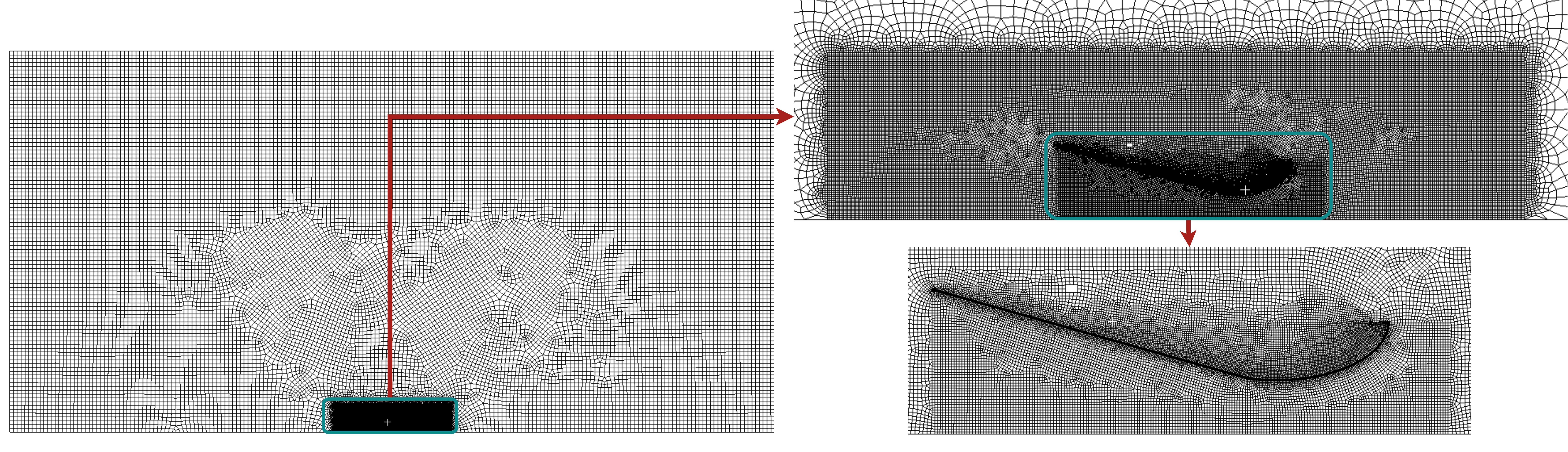}
        \caption{Mesh of the axisymetric geometry with a radius of $12mm$.}       
        \label{fig:mesh} 
\end{figure}

The mesh of the domain is refined near the walls of the geometry to capture the boundary layer and ensure accurate predictions of the flow field as depicted Fig.\ref{fig:mesh}. 
Furthermore, due to the geometric parameters study, the geometry of the BLF and/or its position in the domain may be significantly modified and thus, the mesh needs to be adapted accordingly. 

To ensure that the results are not affected by the mesh size, a convergence study was conducted for the three inlet mass flow rates listed in Table \ref{tab:parameters} (1,10 and 100$g.s^{-1}$) and for two different radii (12 and 30$mm$).The mesh was refined by increasing the number of cells in the region around the BLF. 

It also worth pointing out that the Spalart-Allmaras model has been extended within Ansys Fluent with a $y^+$-insensitive wall treatment, even for the intermediate values in the buffer layer ($1<y^+<30$) \cite{ansys_help}. Consequently, no specific mesh size near the wall is required to capture turbulence effects. 

\begin{figure}[H]
        \centering
        \begin{subfigure}[b]{0.95\textwidth}
            \centering
            \includegraphics[width=0.95\textwidth]{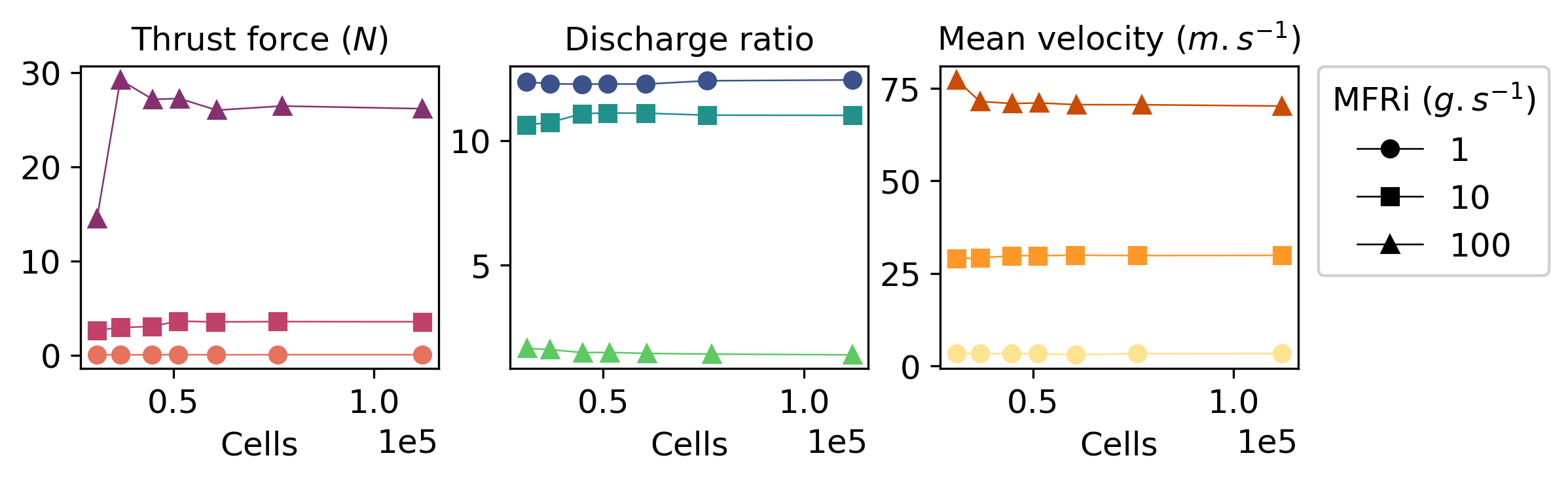}
            \caption{12$mm$ radius}
        \end{subfigure} \\

        \vspace{5px}
        \begin{subfigure}[b]{0.95\textwidth}
            \centering
            \includegraphics[width=0.95\textwidth]{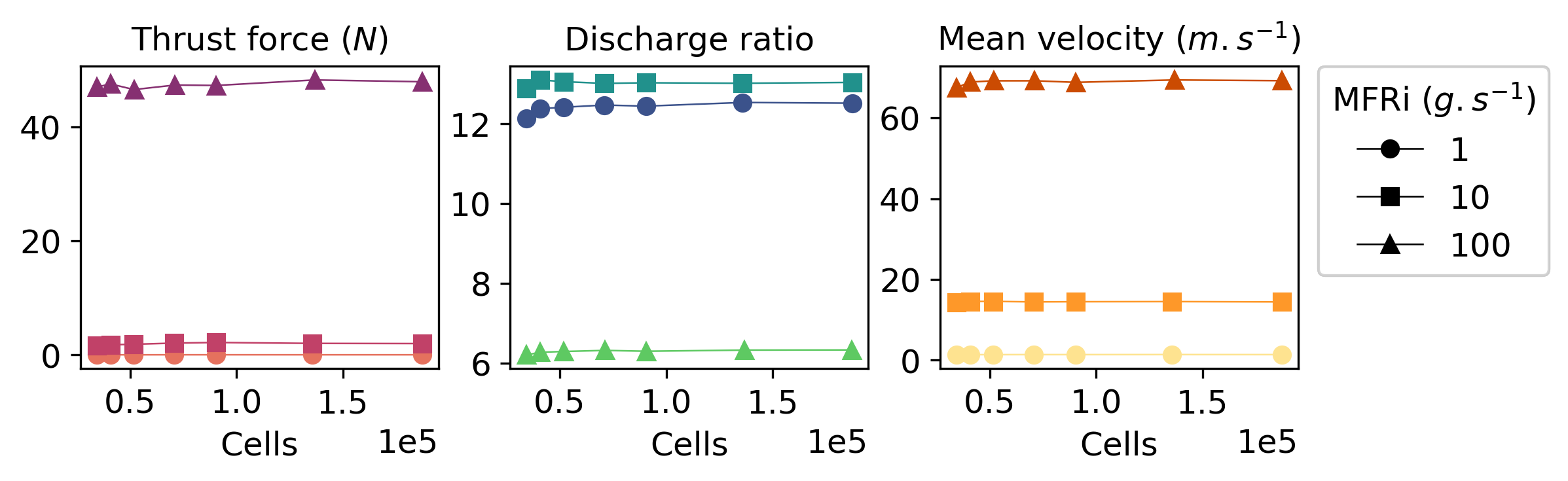}
            \caption{30$mm$ radius}
        \end{subfigure}
        \caption{Convergence of the thrust force, discharge ratio and mean velocity at the BLF outlet for the 3 inlet mass flow rate (MFRi) and for two different radii, 12$mm$ and 30$mm$ respectively.}
        \label{fig:convergence}
    \end{figure}

The results plotted Fig.\ref{fig:convergence} shows that the number of cells for numerical model converges quickly for all the parameters studied. Note also that the number of cells of the model depends on the geometric parameters and more specially of the inner radius with a minimum of $ 7\,10^4$ and a maximum of $ 3.5\,10^5$ cells for radii from $5$ to $200mm$, respectively. 

Finally, the mesh used for the simulations corresponds to the second-to-last refinement level shown in the previous figure.

\section{Results and Discussion}
In this section, we examine the impact of varying slit thickness and radius on the discharge ratio and generated thrust force. These two parameters are tested across three distinct inlet mass flow rates (ranging from $1$ to $100 g.s^{-1}$), five different nozzle sizes (from $0.25$ to $1mm$), and 18 radii (from 2.5 to $200mm$), providing a thorough investigation of the design space.

The tail thickness is arbitrarily set to $5mm$ and the slit located at $20mm$ from the nose.

\subsection{Discharge Ratio}
\begin{figure}[H]
        \centering
                \begin{subfigure}[b]{8cm}
                \includegraphics[trim={0 0 0 11cm},clip, width=6cm]{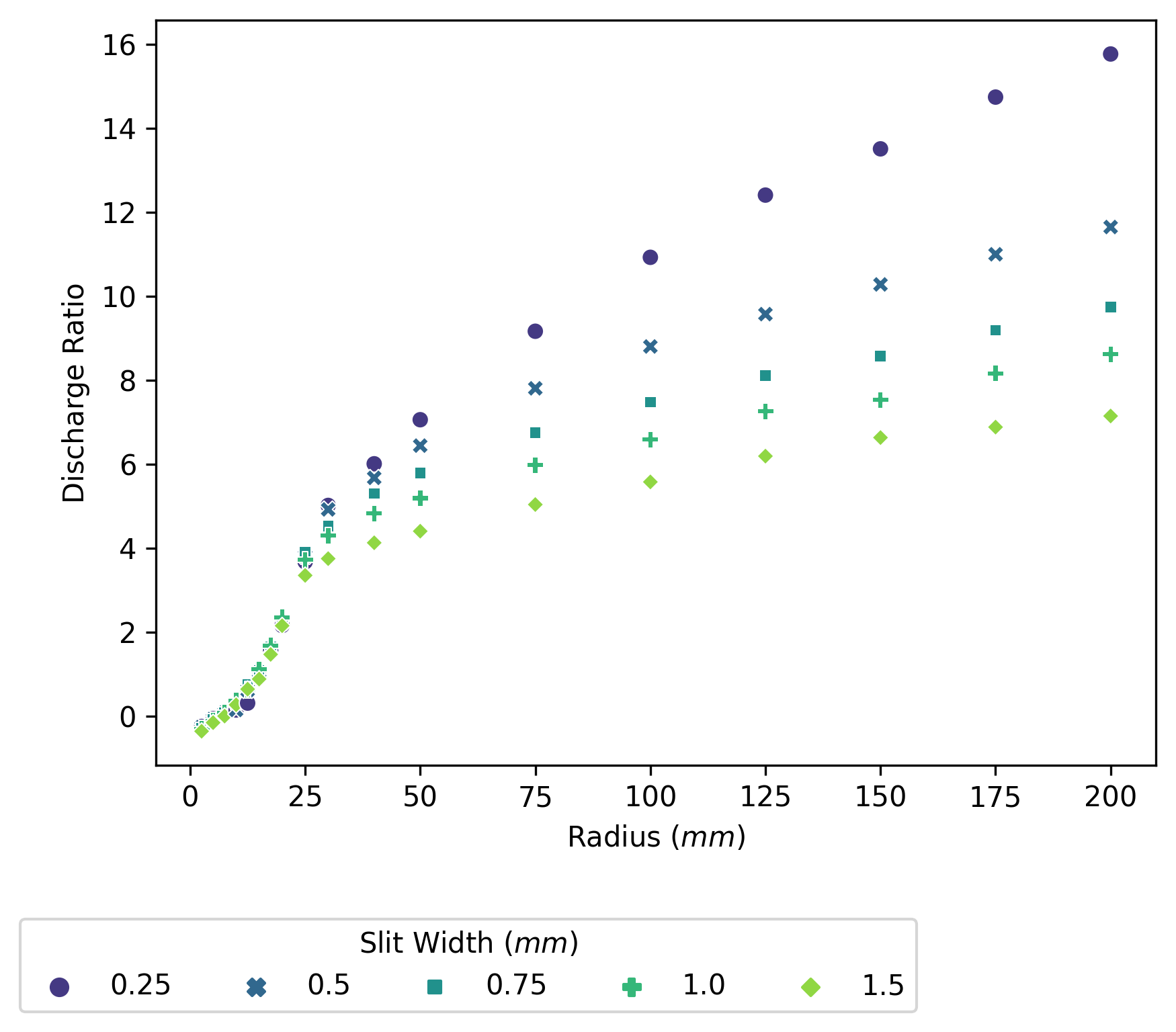}
                \end{subfigure}

                \begin{subfigure}[b]{6cm}
                \includegraphics[height=4.4cm]{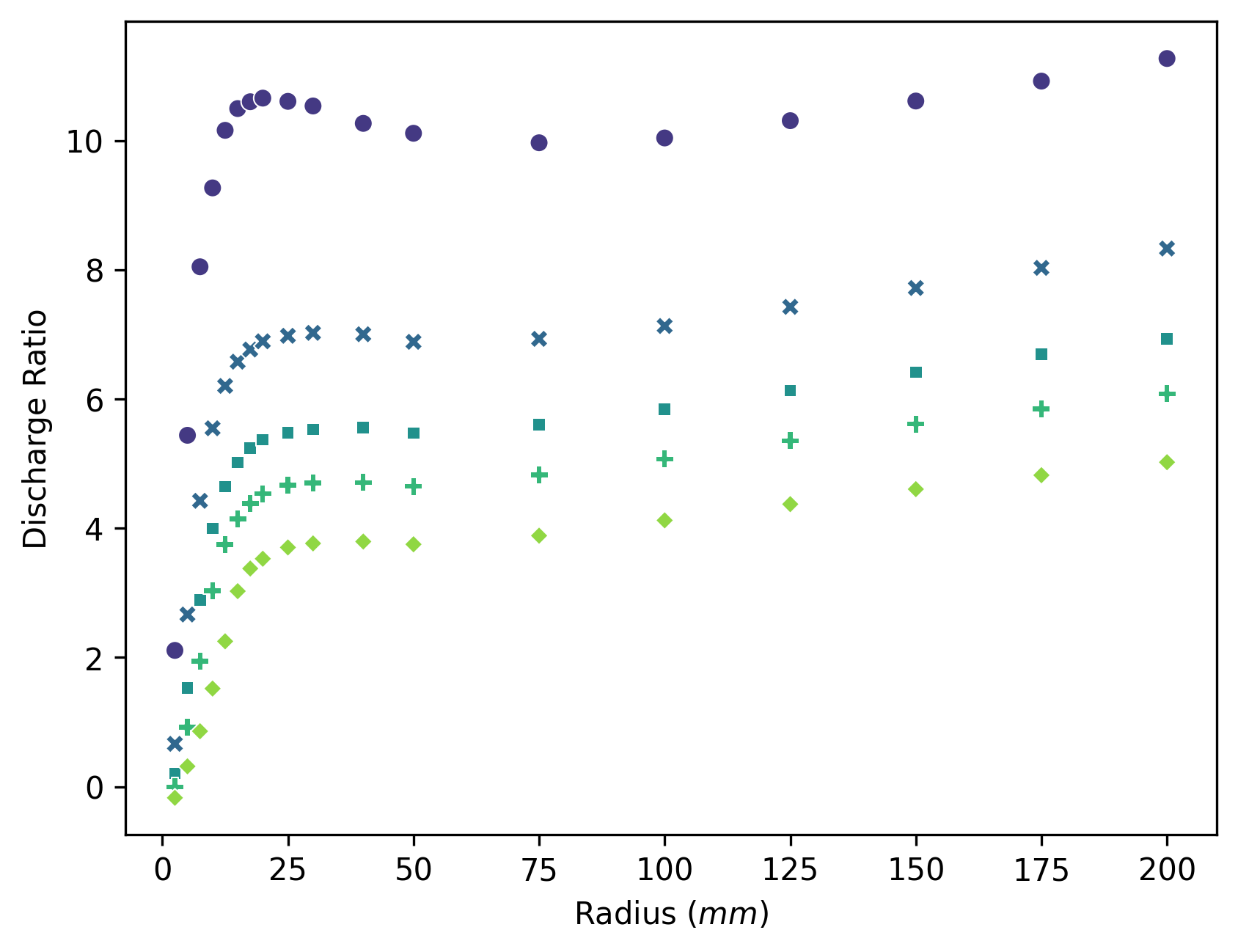}
                \caption{MFRi = 1$g.s^{-1}$}
                \end{subfigure}
                \begin{subfigure}[b]{6cm}
                \includegraphics[height=4.4cm]{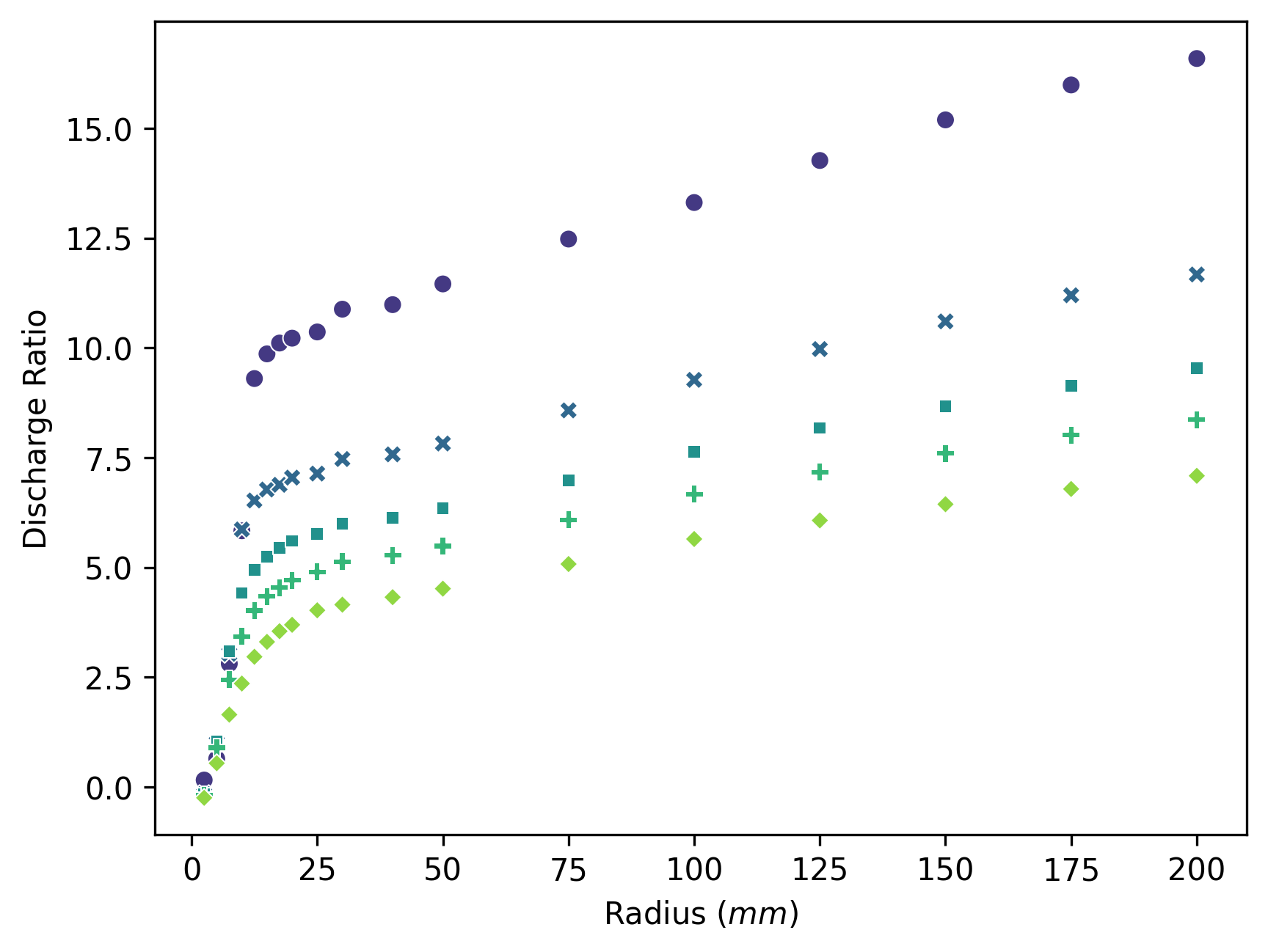}
                \caption{MFRi = 10$g.s^{        -1}$}
                \end{subfigure}
                \begin{subfigure}[b]{6cm}
                \includegraphics[trim={0 2.5cm 0 0cm},clip,height=4.4cm]{data/slit_radius_DR_100.png}
                \caption{MFRi = 100$g.s^{-1}$}
                \end{subfigure}                
        \caption{Discharge ratio for different slit thickness and radius for the three mass flow rates.}
        \label{fig:slitRadiusDR} 
\end{figure}

As previously mentioned in the cited works \cite{jafari2016numerical, jafari2017numerical}, we observe from Fig.\ref{fig:slitRadiusDR} that both the slit thickness and radius significantly influence the fan's discharge ratio and thrust force. For the $1g.s^{-1}$ MFRi, the mass flow rate exhibits a substantial increase at low radii, reaching a local maximum at a radius of $\approx15mm$, which is dependent on the slit thickness, followed by a minor decrease and ultimately displaying a linear increase at higher radii.

For the $10g.s^{-1}$ MFRi, the discharge ratio exhibits a bi-linear trend across all slit thicknesses, characterized by a sharp increase in the discharge ratio at low radii ($<15mm$), and a moderate increase for higher radii.

Regarding the higher MFRi ($100g.s^{-1}$), the discharge ratio presents a tri-linear shape for all slit thicknesses. It exhibits a moderate increase in the discharge ratio at low radii ($<10mm$) with all slits yielding similar discharge ratios. The second regime, between 10 and $30mm$, displays a more pronounced increase where all slits produce coincident results. Finally, the third regime, for radii $>30mm$, showcases a moderate increase in the discharge ratio with the radius, where the influence of the slit thicknesses becomes apparent.

It is important to note that specific combinations of radius and slit thickness result in significant performance improvements. For instance, the discharge ratio increases threefold for all MFRi when the slit thickness is reduced from $1.5mm$ to $0.25mm$.

\subsection{Thrust force}
\begin{figure}[H]
        \centering
        \includegraphics[width=12cm]{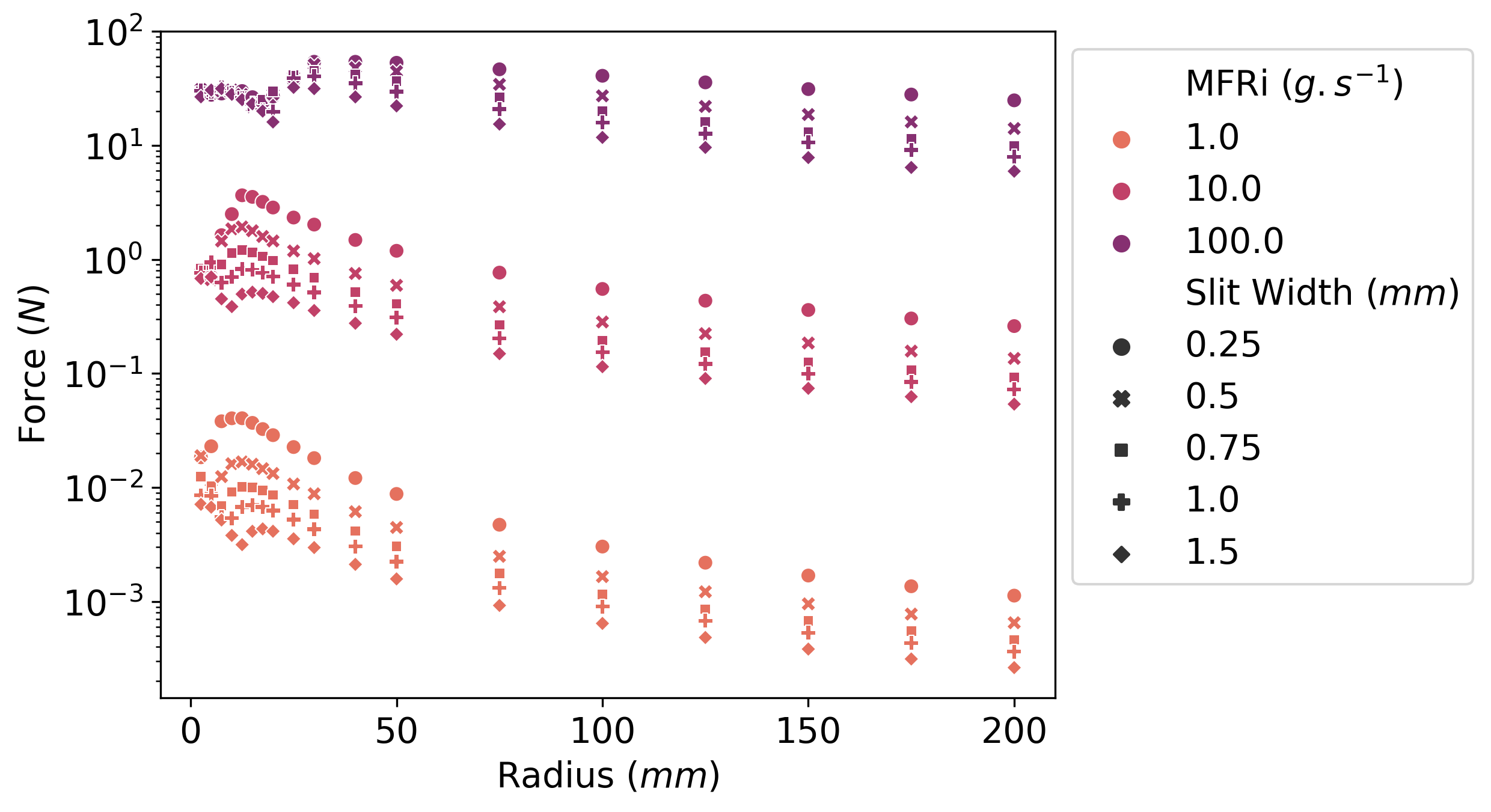}
        \caption{Thrust force variation across different slit thicknesses and radii for the three MFRi.}       
        \label{fig:slitradiusForce} 
\end{figure}

In this section, we examine the thrust force generated by the bladeless fan as it relates to the slit size and radius for the three different MFRi. We observe Fig.\ref{fig:slitradiusForce} that the optimal thrust force is achieved at relatively low radii, specifically between 10 and 30 mm, which varies depending on the chosen slit parameters and MFRi values. 

The impact of the slit size on the generated force is considerably more pronounced than on the discharge ratio, resulting in a five-fold increase in force for MFRi values of 1 and $10g.s^{-1}$. Conversely, for an MFRi of $100g.s^{-1}$, the influence is less noticeable, with a ratio under 2 between the largest and smallest slits, 1.5 and $0.25mm$, respectively.

It is worth noting that the generated thrust forces are not strictly proportional to the MFRi, which is expected, as the minimum slit thickness remains constant across all MFRi values. When examining the maximum forces obtained for each MFRi (using the $0.25mm$ nozzle, as shown in Table \ref{tab:force}), it becomes apparent that the thickness of the slit has a significant impact on the generated thrust force.

\begin{table}[H]
        \centering
        \setlength{\tabcolsep}{2pt}
        \begin{tcolorbox}[tab1,tabularx*={\renewcommand{\arraystretch}{1.25}}{l|c|c},title={Thrust Forces},boxrule=0.8pt, width=7cm]
        \hline
        MFRi ($g.s^{-1}$) &  Force ($N$) &  Ratio($N.s.g^{-1}$)  \\
        \hline
        1  &  0.041 &  0.041 \\
        10 &  3.66 & 0.366 \\
        100 &  54.3 & 0.543 \\
        
        \hline
        \end{tcolorbox}
        \caption{Thrust forces and ratio of thrust force over MFRi for different MFRi with a 0.25$mm$ slit thickness.}
        \label{tab:force}
\end{table}

\subsection{Performance of the BLF geometry}
In the present section, we assess the performance of the bladeless fan both for discharge ratio and thrust force. In the simulations, a mass flow rate was imposed at the inlet, and the pressure inside the bladeless fan adjusted accordingly to meet this boundary condition. As a result, we can evaluate the efficiency of each outcome, whether it is the discharge ratio or the thrust force depending on the pressure inside the BLF. Furthermore, concerning the generated thrust force, one can also compare it to the one generated by a De Laval nozzle \cite{delaval}, which is known to produce high thrust force.
\subsubsection{Discharge ratio}
\begin{figure}[H]
        \centering
                \begin{subfigure}[b]{6cm}
                \includegraphics[trim={0 0 0 11cm},clip, width=6cm]{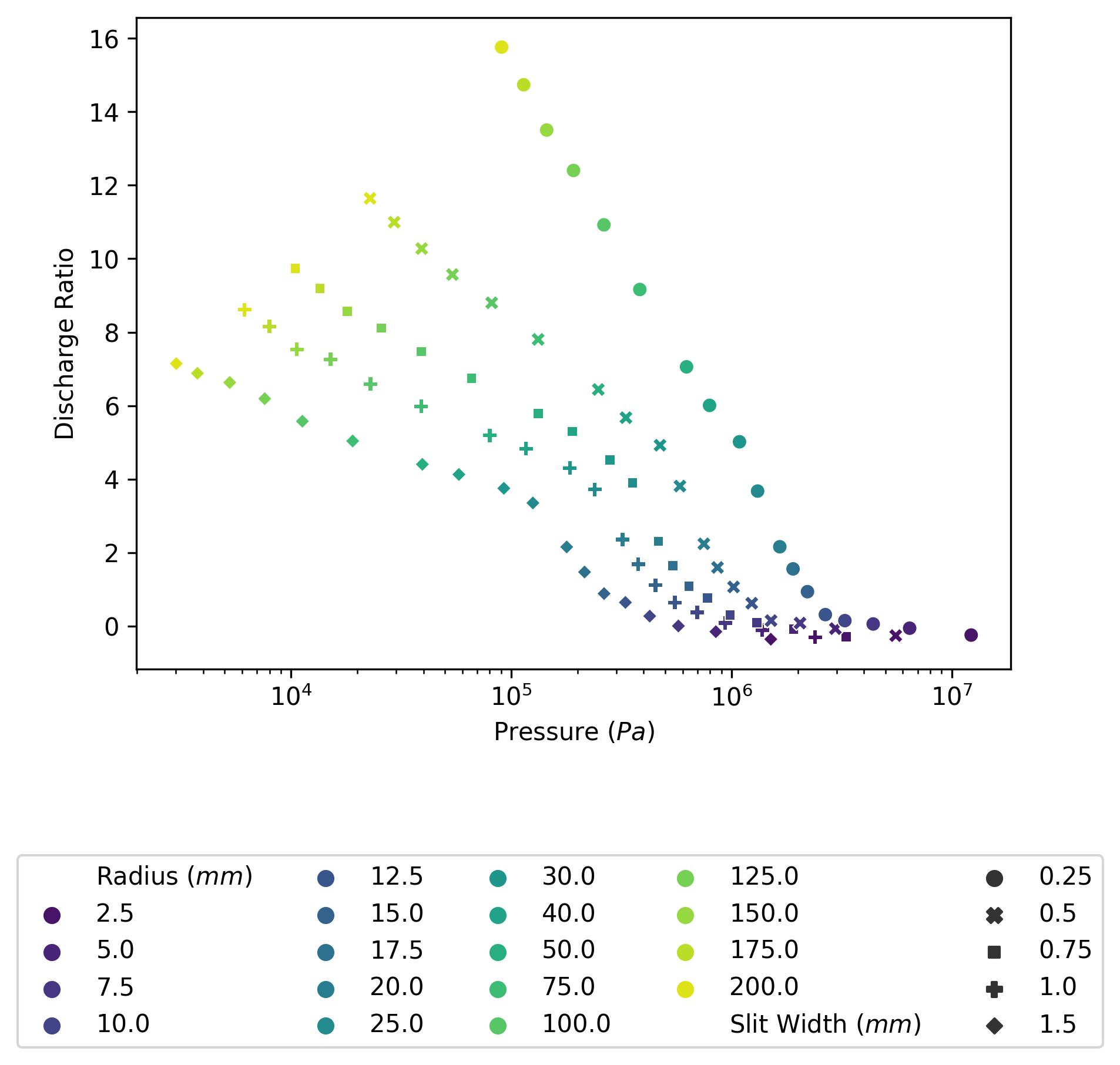}
                \end{subfigure}

                \begin{subfigure}[b]{6cm}
                \includegraphics[height=4.4cm]{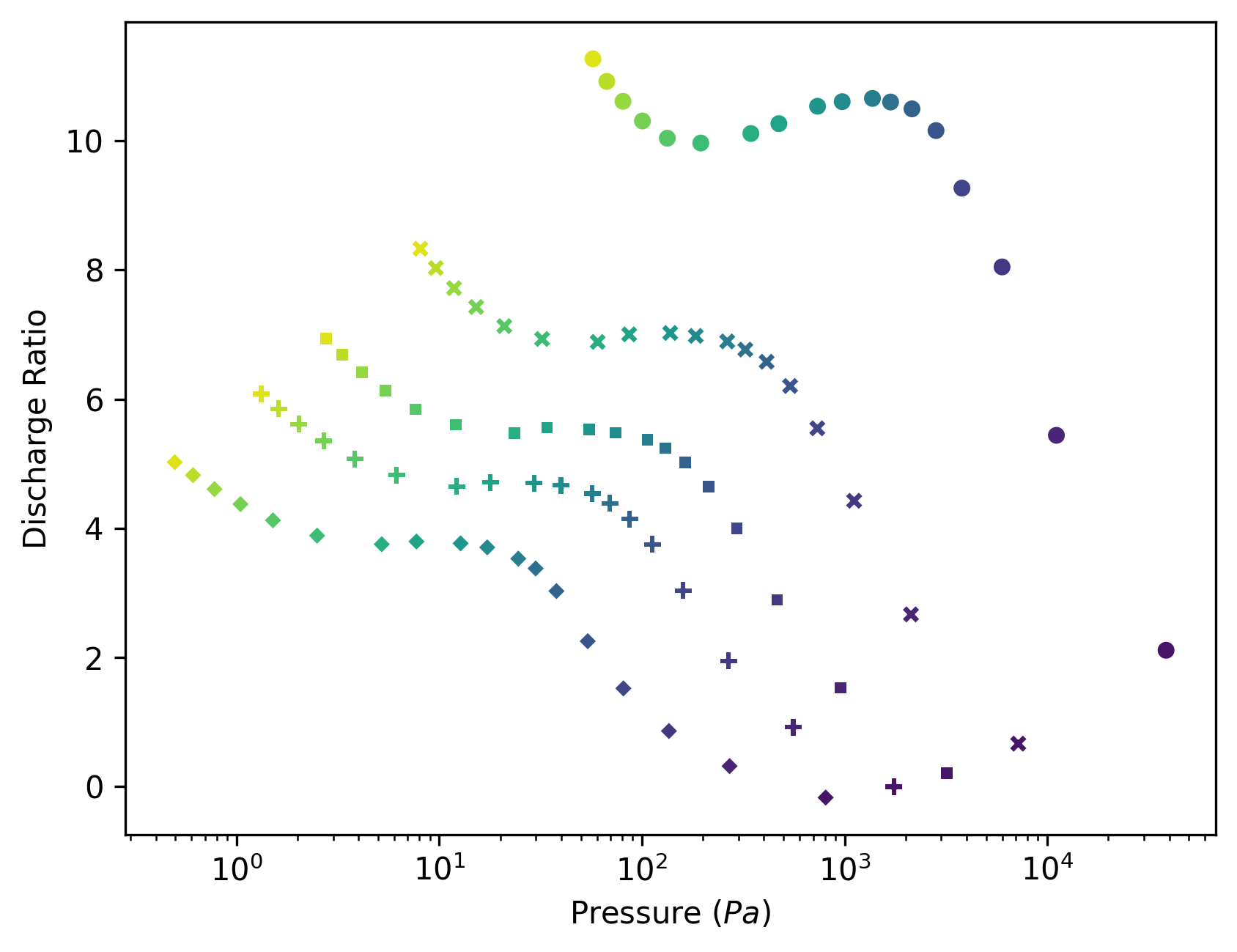}
                \caption{MFRi = 1$g.s^{-1}$}
                \end{subfigure}
                \begin{subfigure}[b]{6cm}
                \includegraphics[height=4.4cm]{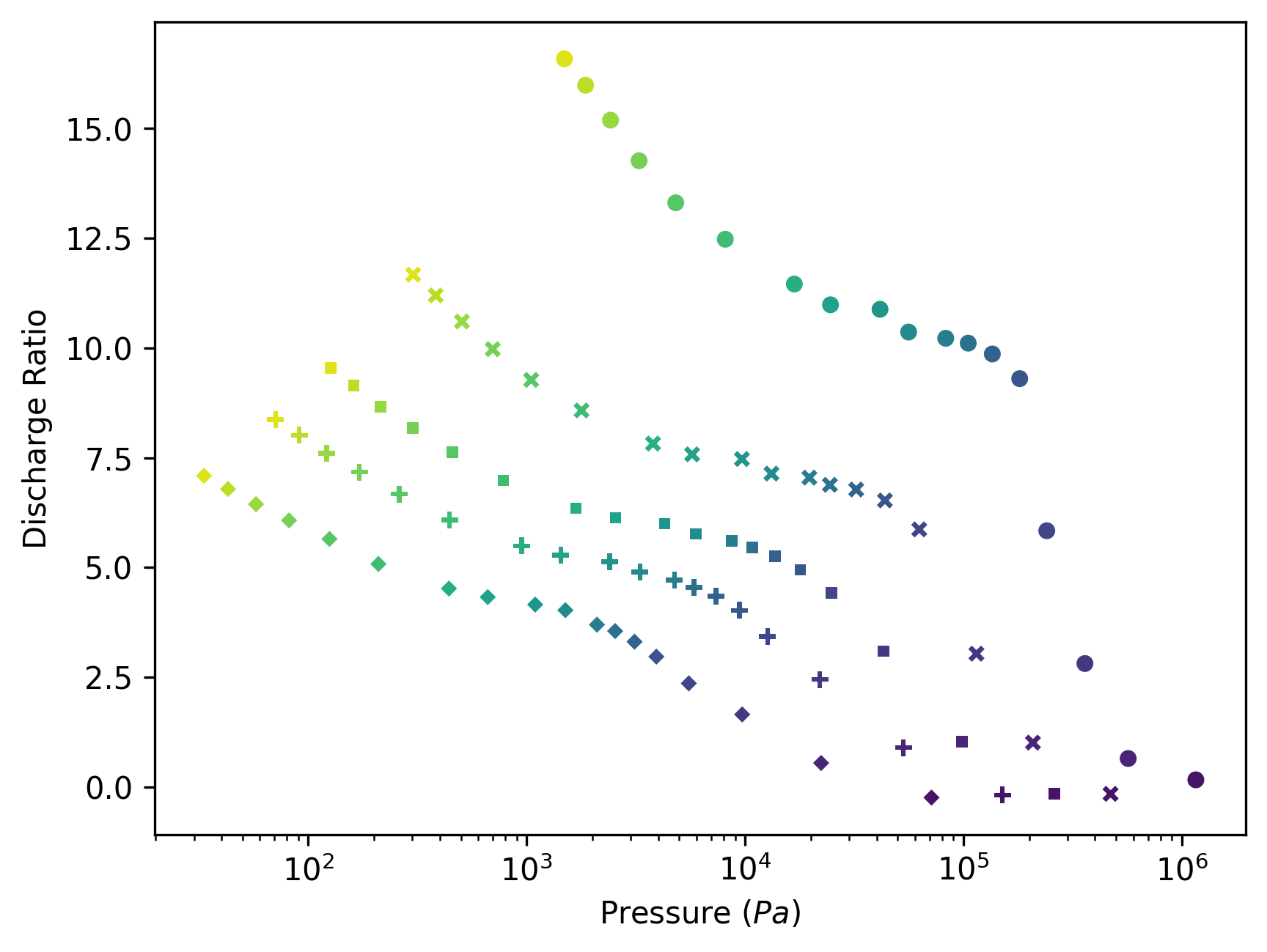}
                \caption{MFRi = 10$g.s^{-1}$}
                \end{subfigure}
                \begin{subfigure}[b]{6cm}
                \includegraphics[trim={0 4.1cm 0 0cm},clip,height=4.4cm]{data/slit_radius_DR_100_efficiency.png}
                \caption{MFRi = 100$g.s^{-1}$}
                \end{subfigure}                
        \caption{Discharge ratio for different slit thicknesses as a function of the pressure inside the BLF, for the three different MFRi values.}
        \label{fig:slitRadiusDRPressure} 
\end{figure}

The analysis depicted in Figure \ref{fig:slitRadiusDRPressure} demonstrates that there is no linear correlation between the discharge ratio and the pressure inside the bladeless fan. Moreover, the presence of local optima suggests that the discharge ratio is influenced by factors beyond the pressure alone. Additionally, it is noteworthy that the more efficient designs, which demand lower pressure, exhibit larger radii. This trend holds true regardless of the slit thickness or mass flow rate inlet (MFRi) under consideration.

\subsubsection{Thrust force}
While the thrust force generated by a bladeless fan (BLF) may be considered weak, this geometry offers several advantages that make it a suitable choice for some applications. One key benefit of the BLF is its compact size, which allows for easy integration into systems. Additionally, since the BLF is bladeless, it does not require any moving parts, leading to reduced maintenance costs and a lower risk of failure. Finally, the BLF is a silent device, which is highly desirable for applications where noise reduction is important and some companies, such as Jetoptera \cite{jetoptera}, leverage similar geometries or systems to develop airborne vehicles.

\begin{figure}[H]
        \centering
        \includegraphics[width=12cm]{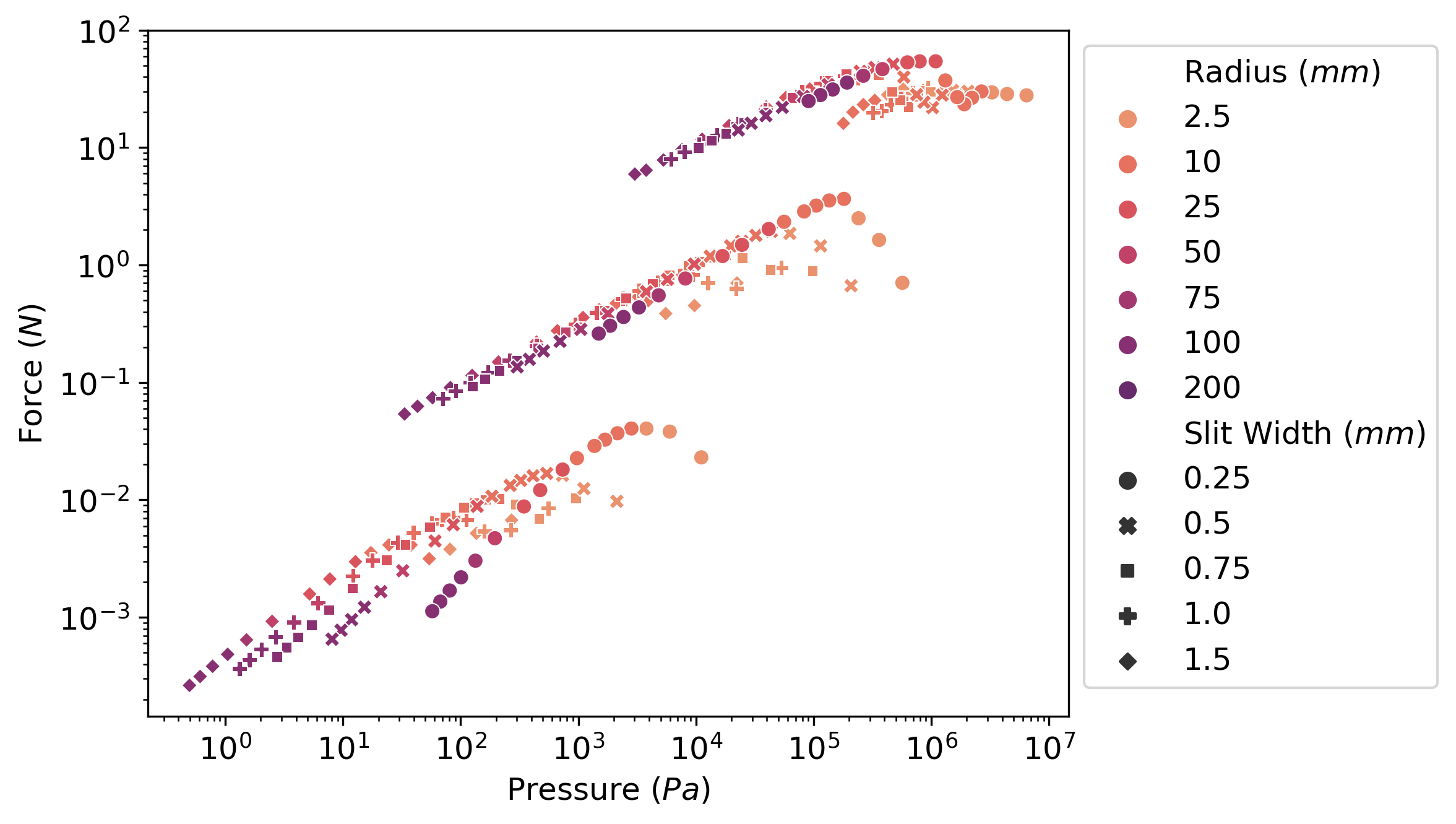}        
        \caption{Generated thrust force for different slit thicknesses and radii, as a function of the pressure inside the BLF.}
        \label{fig:slitRadiusForcePressure} 
\end{figure}

Figure \ref{fig:slitRadiusForcePressure} depicts the generated thrust force as a function of the pressure in the BLF. As previously observed Fig. \ref{fig:slitradiusForce}, the generated forces reach a maximum when decreasing the radius, and then decrease for each slit thickness. As one can expect, the increase in force is accompanied by an increase in the pressure inside the BLF for each slit thickness. However, once the maximum force is reached and begins to decrease, the required pressure continues to increase, entering a regime that is significantly inefficient. 

 \subsubsection*{De Laval thrust force}
For compressed air, a converging-diverging nozzle, the so-called \textit{De Laval} nozzle \cite{delaval}, is generally considered the most effective nozzle in terms of thrust force generation. This is because a converging-diverging nozzle can efficiently convert the pressure energy of the compressed air into kinetic energy, resulting in a high-velocity (eventually supersonic) jet of air. 

Newton's third law of motion can be applied to determine the force exerted by the expelled gas, which is commonly referred to as thrust in the field of aerodynamics. The thrust generated by the exhaust gases can be calculated using the mass flow rate and the exit velocity at the nozzle exit. The relationship between thrust, mass flow rate, and exit velocity at nozzle exit is expressed as:

\begin{equation}
        F = v_e \dot{m}
        \label{eqForce}
\end{equation}

where $F$ is force exerted or thrust, $\dot{m}$ the mass flow rate of the expelled gas and $v_e$ the exit velocity of the gas at the nozzle exit. In the case of a a de Laval nozzle, the fluid velocity can surpass the speed of sound (i.e., Mach 1), allowing for the maximization of thrust force. The exit velocity can be analytically calculated using Equation \ref{eqPoussee} \cite{sutton2016rocket}

\begin{equation}
        {\displaystyle v_{e}={\sqrt {2C_{p}T_{0}\left[1-\left({\frac {P_{e}}{P_{0}}}\right)^{\frac {\gamma -1}{\gamma }}\right]}}={\sqrt {{\frac {T_{0}R}{M}}\cdot {\frac {2\gamma }{\gamma -1}}\cdot \left[1-\left({\frac {P_{e}}{P_{0}}}\right)^{\frac {\gamma -1}{\gamma }}\right]}}}
        \label{eqPoussee}
\end{equation}
where :
\begin{align*}
        v_e &: \text{Velocity of the gas at the nozzle exit (m/s)} \\
        T_0 &: \text{Total temperature at the nozzle inlet (K)} \\
        R &: \text{Universal gas constant for ideal gases} \\
        M &: \text{Molar mass of the gas} \\
        \gamma &: \text{Adiabatic index} \\
        C_p &: \text{Specific heat capacity of the gas at constant pressure} \\
        P_e &: \text{Static pressure of the gas at the nozzle exit} \\
        P_0 &: \text{Total pressure of the gas at the nozzle inlet}
\end{align*}

\begin{figure}[H]
        \centering
        \includegraphics[width=12cm]{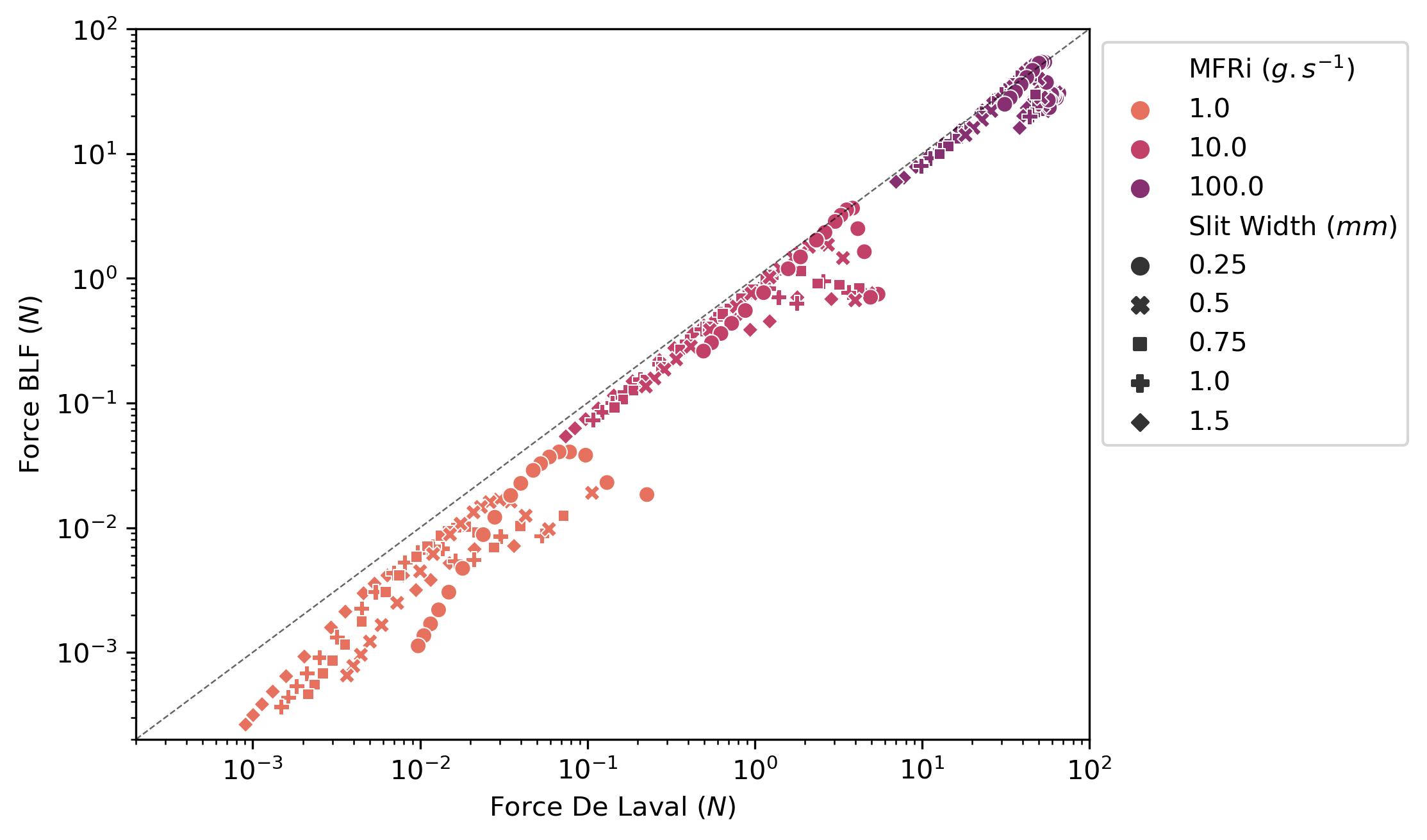}
        \caption{Thrust force from the BLF as a function of the thrust force from De Laval nozzle across different slit thicknesses and radii for the three MFRi.}       
        \label{fig:slitradiusForceEfficiency} 
\end{figure}

Using the pressure obtained in simulations for calculating $v_e$, it is possible to compare the thrust force generated by the BLF with the thrust force generated by a De Laval nozzle. The results are presented in Fig. \ref{fig:slitradiusForceEfficiency}. When the points are located in the lower-right corner, the force generated by the BLF is less than that generated by the De Laval nozzle. Additionally, the closer the data points are to the dotted line representing the first bisector, the more energetically efficient the generated forces are

We also observe that for certain configurations, the thrust force generated by the BLF is of the same order of magnitude as that of the De Laval nozzle when the input pressure and MFRi are identical: some points of the BLF-generated thrust force have the same abscissa (and thus the same internal pressure in the geometry) for different values of thrust force, indicating that certain configurations of the BLF are more efficient than others for generating thrust forces.

\begin{figure}[H]
        \centering
        \includegraphics[width=12cm]{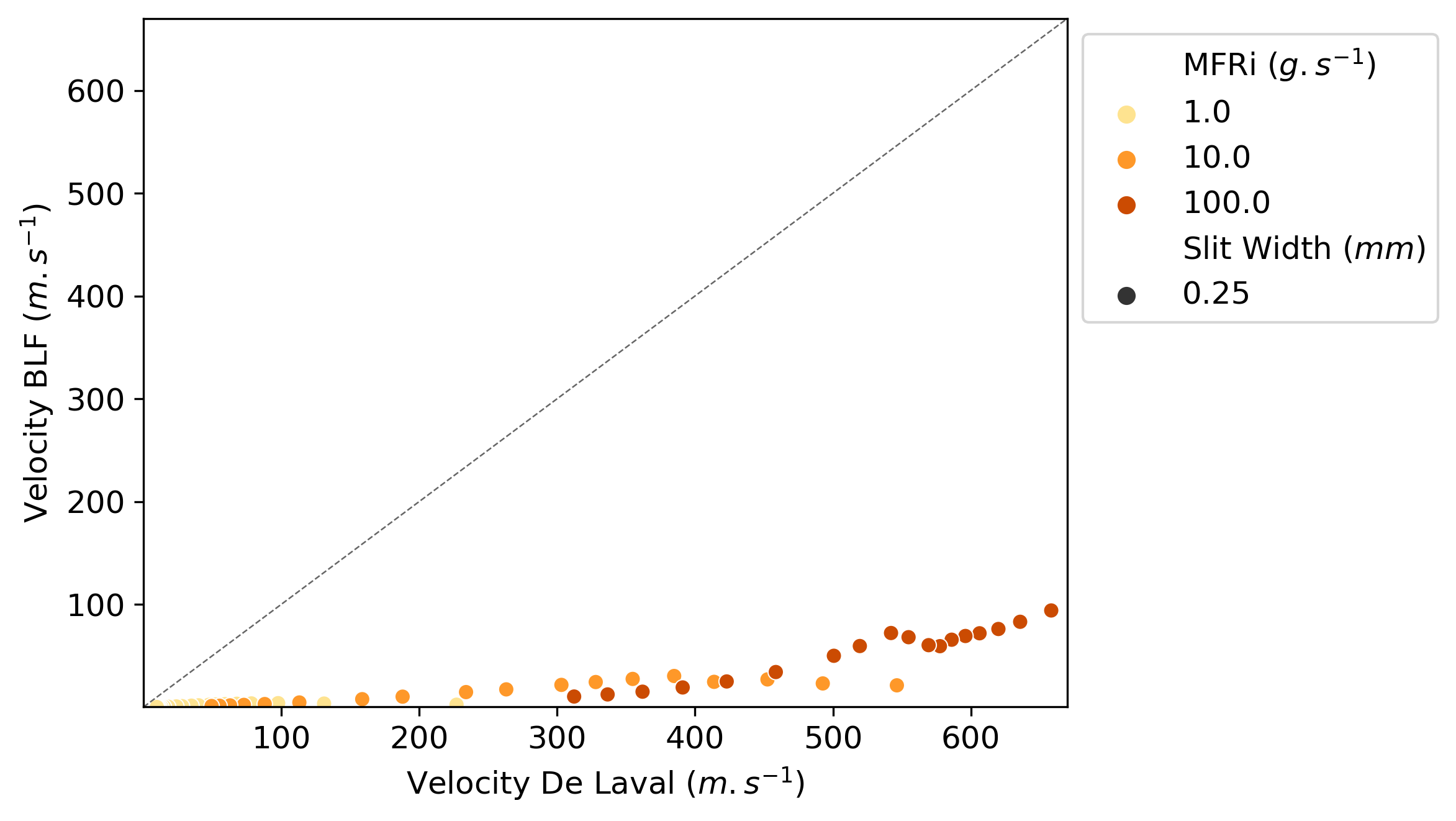}
        \caption{Outlet velocities from the BLF as a function of the $v_e$ from De Laval nozzle for the three MFRi and for a slit thickness of $0.25mm$.}       
        \label{fig:slitradiusForceEfficiencyVel} 
\end{figure}

Moreover, upon comparing the outlet velocities $v_e$ of the bladeless fan (BLF) with a $0.25mm$ slit to those of the equivalent De Laval nozzle (see Fig. \ref{fig:slitradiusForceEfficiencyVel}), it is evident that while the BLF generates comparable thrust, the outlet velocities are significantly lower. This finding further supports the notion that the bladeless fan geometry offers a significant reduction in noise compared to the equivalent convergent-divergent nozzle \cite{lighthill1952sound, jafari2017numerical}.

The velocity reduction for the BLF geometry is compensated by the large increase mass flow rate, i.e. corresponding to the high discharge ratio observed previously, to generate equivalent thrust force. This  phenomenon can be a real asset as the gas from the BFL geometry is mixed with the ambient air which, in turn, will widely mitigate its temperature either by warming (in case of compressed air decompression for instance) or cooling it (for exhaust gas).


\section{Conclusion}
This study examined the impact of slit thickness and radius on the performance of the BladeLess Fan geometry with compressed air across a wide range of inlet mass flow rates slit thickness and radii, focusing on low to moderate radii. 

The findings demonstrate that at lower radii, both the discharge ratio and generated thrust have a nonlinear relationship with these parameters, and an optimal radius is required to achieve maximum thrust force with specific combinations leading to substantial improvements in performance. For higher radii, and in accordance with the literature, the discharge ratio show a linear increase with the radius while the thrust force decreases.
It has also been found that a discharge ratio greater than 10 can be achieved with combinations of small slit thickness and radius. 

Moreover, the thrust force generated by the bladeless fan for a specific geometry is in the same order of magnitude as that of a De Laval nozzle, with additional benefits of reduced noise and mixing of outlet gas with surrounding air.

The numerical simulations conducted using Ansys Fluent provide a better understanding of the performance characteristics of bladeless fans, which can inform the design and optimization of similar devices. Indeed for a given pressure in the BLF body, performance can vary widely depending on the choice of those geometric parameters as shown in last section of the article.

Overall, this study highlights the importance of carefully considering a range of geometric parameters in the design and optimization of bladeless fans. The results provide valuable insights that can inform future research in this field, and aid in the development of more efficient and effective bladeless fan designs.

\authorcontributions{
	Conceptualization, Guillaume Maîtrejean, Mohamed Karrouch and Didier Bleses; Methodology, Guillaume Maîtrejean; Project administration, Guillaume Maîtrejean, François Truong and Denis Roux; Resources, Guillaume Maîtrejean; Software, Guillaume Maîtrejean, Lucas Antoniali, Merlin Kempf and Emma Ferreira-lopes; Supervision, Guillaume Maîtrejean; Writing – original draft, Guillaume Maîtrejean; Writing – review \& editing, Guillaume Maîtrejean.
All authors have read and agreed to the published version of the manuscript.}

\funding{LRP is part of the LabEx Tec21 (ANR-11-LABX-0030) and of the PolyNat Carnot Institute (ANR-11-CARN-007-01).}

\dataavailability{The data presented in this study are available on request from the corresponding author.} 



\conflictsofinterest{The authors declare no conflict of interest.} 





%

\begin{adjustwidth}{-\extralength}{0cm}

\reftitle{References}


\bibliography{biblio}

\PublishersNote{}
\end{adjustwidth}
\end{document}